\documentclass{revtex4}

\begin{document}
\sloppy

\title{Nambu-Hamiltonian flows associated with discrete maps}
\author{Satoru SAITO}
\email[email : ]{saito@phys.metro-u.ac.jp}
\affiliation{Department of Physics, Tokyo Metropolitan University,\\
Minamiohsawa 1-1, Hachiohji, Tokyo 192-0397 Japan}
\author{Akira SHUDO}
\email[email : ]{shudo@phys.metro-u.ac.jp}
\affiliation{Department of Physics, Tokyo Metropolitan University,\\
Minamiohsawa 1-1, Hachiohji, Tokyo 192-0397 Japan}
\author{Jun-ichi YAMAMOTO}
\email[email : ]{yjunichi@kiso.phys.metro-u.ac.jp}
\affiliation{Department of Physics, Tokyo Metropolitan University,\\
Minamiohsawa 1-1, Hachiohji, Tokyo 192-0397 Japan}
\author{Katsuhiko YOSHIDA}
\email[email : ]{yoshida@kiso.phys.metro-u.ac.jp}
\affiliation{School of Science, Kitasato University,\\ 1-15-1 Kitasato Sagamihara, Kanagawa, 228-8555 Japan}
\begin{abstract}
For a differentiable map $(x_1,x_2,\cdots, x_n)\rightarrow (X_1,X_2,\cdots, X_n)$ that has an inverse, we show that there exists a Nambu-Hamiltonian flow in which one of the initial value, say $x_n$, of the map plays the role of time variable while the others remain fixed. We present various examples which exhibit the map-flow correspondence.
\end{abstract}

\maketitle
\baselineskip 20pt
\section{INTRODUCTION}

In our previous paper\cite{SSYY} we have shown that, for a given map $(x,y)\rightarrow (X,Y)$, there exists a Hamiltonian flow in which $y$ plays the role of time variable, {\it iff} the map is differentiable and has an inverse. We would like to present various examples in this paper in order to clarify the correspondence. We also show that our scheme can be generalized naturally to Nambu-Hamiltonian flow\cite{Nambu,Takhtajan} when the dimension of the map is greater than two.

Before starting our discussion, it will be worth-while to explain the motivation of our study. Let $\mbox{\boldmath$x$}$ be a set of variables either real or complex, and $f(\mbox{\boldmath$x$})$ a differentiable function. For a given $f$ we can consider a sequence of the map
\begin{equation}
\mbox{\boldmath$x$} \rightarrow f(\mbox{\boldmath$x$}) \rightarrow f^{(2)}(\mbox{\boldmath$x$}) \rightarrow \cdots \rightarrow f^{(m)}(\mbox{\boldmath$x$}),\qquad \left(f^{(2)}:=f(f(\mbox{\boldmath$x$})),\quad {\rm etc.}\right).
\label{map}
\end{equation}
We call this sequence a dynamical system generated by $f$. 

We encounter such systems in various occasion in mathematics, physics and in the nature. All of complex dynamical 
systems are examples in mathematics. B\"acklund transformations which characterize integrable systems, connection 
formulae of Stokes geometry are other examples. Renormalization group formula is an important example of dynamical 
systems in physics.

When a set of initial values $\mbox{\boldmath$x$}$ is given, the map (\ref{map}) determines values of them after $m$ steps. From this information, 
however, it is not obvious if one can predict the values for other initial values. It will be desirable if 
there is a systematic method to analyze the initial value dependence of the map.

In order to answer this question, we have shown, in the previous paper\cite{SSYY}, the following proposition

\noindent
{\bf Proposition 1}
{\it
Let a differentiable and invertible map $(x,y)\rightarrow (X,Y)$ of the plane be given and $H=H(X,Y)$ be the function given by
\begin{equation}
H(X,Y)=\int^{x(X,Y)}(\det J)dx,
\label{Hamiltonian}
\end{equation}
in which 
$$
\det J={\partial(X,Y)\over\partial (x,y)}
$$
is the Jacobian of the map satisfying
$$
{\partial (\det J)\over\partial y}=0,
$$
then the following system of Hamilton's equations hold:
\begin{equation}
{dX\over dy}=-\ {\partial H\over\partial Y},\qquad 
{dY\over dy}= {\partial H\over\partial X}.
\label{Hamilton's equations}
\end{equation}
}

Note that the initial value $y$ of the map plays the role of time variable. The Hamiltonian is determined if the Jacobian of the map is given as a function of $x$ and $x$ is known as a function of $X$ and $Y$. The dependence of $(X,Y)$ on the initial value $y$ is determined by solving the Hamilton equations (\ref{Hamilton's equations}). In the case of $\det J=1$ the area is preserverd through the map and the Hamiltonian turns out to be $x(X,Y)$ itself.

The purpose of this paper is to generalize our scheme to higher dimensional maps. We will derive naturally a Nambu-Hamiltonian flow with $n-1$ Hamiltonians from an $n$ dimensional map. Another purpose of this paper is to present various examples to show how they fit to our scheme. 

If the initial value of one of the variables is changed the map will determine another set of final values. As far as the map is fixed, the change of the final values are known as well. One might think that since the map is known, there is no use of the Nambu-Hamilton equations. It is, however, not obvious if there exists a universal formula which describes the change of the final data of an arbitrary map. What we claim is that the Hamilton equation describes the change in the case of any two dimensional map, whereas the Nambu equation plays the role of universal formula in the case of higher dimensional map. 

In fact solutions to the Nambu-Hamilton equations discussed in this paper will not give any new information, but simply reproduce the map under certain initial conditions which must be chosen carefully. Nevertheless we believe that the use of our scheme will be valuable in the sense that it provides a different point of view of the map. The situation is similar to the relation between the Hamilton Jacobi equations and Hamilton's equations of motion. What we emphasize here is the correspondence between two schemes, namely, the map and the Nambu-Hamiltonian flow.

The generalization of Proposition 1 to the higher dimensional map will be discussed in \S 2. Some implications of the map-flow correspondence will be discussed in \S 3. In \S4 we present four examples of two and three dimensional maps. The correspondence between the recurrence formula and the differential equation for the Hermite polynomials offers an example of two dimensional map-flow correspondence. The H\'enon map, which is a canonical form of the two dimensional polynomial diffeomorphism, will be studied as an example of nonintegrable systems. The 3-point KdV map will provide an example of three dimensional map. Since this is a completely integrable system, we can reduce the number of independent dynamical variables from three to two and obtain a two dimensional flow instead of three dimensional Nambu flow. We will show that the two dimensional view of the flow becomes quite complicated compared with three dimensional one. Another example of three dimensional map is the $q$-difference 3-point Painlev\'e IV map. The integrability of this map is guaranteed either in the continuous limit of the map or for a particular values of the set of parameters.

Although the generalization to higher dimensional map is straightforward, one dimensional map must be discussed separately. We will show in appendix how it can be described in terms of Hamiltonian flow.

\section{Nambu-Hamiltonian flows}

In this section we generalize our scheme of deriving Hamiltonian systems associated with two dimensional map to higher dimensional ones. Since we naturally obtain the Nambu-Hamiltonian systems\cite{Nambu,Takhtajan} it will be convenient to summarize notations briefly before we start our discussion. 

For an $n$ dimensional map 
\begin{equation}
(x_1,x_2,\cdots, x_n) \ \rightarrow \ (f_1,f_2,\cdots,f_n)
\label{n-dim map}
\end{equation}
we adopt, in this paper, the definition of the Nambu bracket\cite{Nambu} $\{f_1,f_2,\cdots,f_n\}$ given by the Jacobian of the map
$$
\{f_1,f_2,\cdots,f_n\}:={\partial(f_1,f_2,\cdots,f_n)\over\partial(x_1,x_2,\cdots, x_n)}.
$$
Note that the following identity holds for arbitrary function $g$ of $(x_1,x_2,\cdots, x_n)$:
\begin{equation}
\sum_{l=1}^n\{f_1,f_2,\cdots,f_{j-1},x_l,f_{j+1},\cdots, f_n\}{\partial g\over\partial x_l}=\{f_1,f_2,\cdots,f_{j-1},g,f_{j+1},\cdots, f_n\}.
\label{identity}
\end{equation}

If $(X_1,X_2,\cdots,X_n)$ are $n$ dynamical variables, the Nambu equations of motion\cite{Nambu,Takhtajan} are
$$
{dX_j\over dt}={\partial(H_1,H_2,\cdots,H_{n-1},X_j)\over \partial(X_1,X_2,\cdots,X_{n})},\qquad j=1,2,\cdots,n
$$
where $H_j,\ j=1,2,\cdots,n-1$ are Hamiltonians. The equation of motion of a function $f=f(X_1,X_2,\cdots,X_n)$ follows, using (\ref{identity}), as
\begin{eqnarray*}
{df\over dt}&=&
\sum_{j=1}^n{dX_j\over dt}{\partial f\over\partial X_j}
=\sum_{j=1}^n{\partial(H_1,H_2,\cdots,H_{n-1},X_j)\over \partial(X_1,X_2,\cdots,X_{n})}{\partial f\over\partial X_j}\\
&=&{\partial(H_1,H_2,\cdots,H_{n-1},f)\over \partial(X_1,X_2,\cdots,X_{n})}.
\end{eqnarray*}
By definition of the Nambu bracket we find
$$
{dH_j\over dt}=0,\qquad j=1,2,\cdots,n-1.
$$

Our question is whether there exists a Nambu-Hamiltonian flow associated with the $n$ dimensional map. The answer is given by the following proposition.

\noindent
{\bf Proposition 2}

{\it 
Let a differentiable and invertible map 
\begin{equation}
\mbox{\boldmath$x$}=(x_1,x_2,\cdots,x_n)\quad\rightarrow\quad \mbox{\boldmath$X$}=(X_1,X_2,\cdots,X_n),
\label{map x-X}
\end{equation}
be given and $H_j,\ j=1,2,\cdots,n-1$ be functions of $(X_1,X_2,\cdots,X_n)$ defined by
\begin{equation}
H_j=x_j(\mbox{\boldmath$X$}),\quad (j=1,2,\cdots, n-2),\qquad 
H_{n-1}
=\int^{x_{n-1}(\mbox{\boldmath$X$})}(\det J)\ dx_{n-1},
\label{H's}
\end{equation}
in which
$$
\det J={\partial(X_1,X_2,\cdots,X_n)\over\partial(x_1,x_2,\cdots, x_n)}
$$
is the Jacobian of the map satisfying
$$
{\partial(\det J)\over\partial x_n}=0.
$$
Then the following system of Nambu-Hamilton's equations hold:
\begin{equation}
{dX_j\over dx_n}={\partial(H_1,H_2,\cdots,H_{n-1},X_j)\over \partial(X_1,X_2,\cdots, X_n)},\qquad 
j=1,2,\cdots,n.
\label{Nambu-Hamilton eq}
\end{equation}
}
\vglue 0.5cm

\noindent
{\bf Remarks}
\begin{enumerate}
\item
The previous proposition 1 is a special case of $n=2$.
\item
The initial value $x_n$ of the map (\ref{map x-X}) plays the role of time variable of the Nambu-Hamiltonian flow. Any other choice of the time variable among $x_j$'s generates a corresponding flow.
\item
By solving the set of Nambu-Hamilton's equations (\ref{Nambu-Hamilton eq}) we can reconstruct the map $(H_1,H_2,\cdots$ $, H_{n-1},x_n)\rightarrow (X_1,X_2,\cdots, X_n)$ only if we choose initial conditions properly. If we consider all Nambu-Hamilton's equations associated with all possible choices of the `time' variable the map is totally restored.
\end{enumerate}

\noindent
{\bf Proof}

We want to have Hamiltonians $H_j$'s which do not depend on $x_n$. This is true, by definition of the Hamiltonian's (\ref{H's}), in the cases of $j=1,2,\cdots, n-2$, {\it i.e.}, 
\begin{equation}
{dH_j\over dx_n}={dx_j\over dx_n}=0,\qquad j=1,2,\cdots,n-2,
\label{dH_j/dx_n=0}
\end{equation}
as long as $x_1,x_2,\cdots, x_{n-2}$ are independent on $x_n$. This is not true in the case of $x_{n-1}$, since $\det J$, hence $H_{n-1}$, depends on $x_n$ in general. In order to eliminate the $x_n$ dependence of $H_{n-1}$, we relate $x_{n-1}$ with $x_n$ so that the net dependence of $H_{n-1}$ on $x_n$ turns neutral. 

From the expression (\ref{H's}) of the Hamiltonian $H_{n-1}$, we first notice the relations
\begin{eqnarray*}
{dH_{n-1}\over dx_n}&=&{dx_{n-1}\over dx_n}\det J+\int^{x_{n-1}}{\partial\det J\over\partial x_n}\ dx_{n-1}\\
{\partial H_{n-1}\over \partial x_n}&=&{\partial x_{n-1}\over \partial x_n}\det J+\int^{x_{n-1}}{\partial\det J\over\partial x_n}\ dx_{n-1}.
\end{eqnarray*}
Using these relations we can derive the following:
\begin{eqnarray*}
{\partial(H_1,H_2,\cdots,H_{n-1},x_{n-1})\over\partial(x_1,x_2,\cdots,x_n)}&=&\left|
\begin{array}{cc}\displaystyle{{\partial H_{n-1}\over\partial x_{n-1}}}&\displaystyle{{\partial H_{n-1}\over\partial x_{n}}}\\
1&\displaystyle{{\partial x_{n-1}\over\partial x_n}}\\
\end{array}\right|\\
&=&
\det J{\partial x_{n-1}\over\partial x_n}-{\partial H_{n-1}\over\partial x_n}\\
&=&
-\int^{x_{n-1}}{\partial \det J\over\partial x_n}dx_{n-1}\\
&=&{dx_{n-1}\over dx_n}\det J-{dH_{n-1}\over dx_{n}}.
\end{eqnarray*}
Therefore $dH_{n-1}/dx_n=0$ is satisfied iff 
$$
{dx_{n-1}\over dx_n}={1\over\det J}{\partial(H_1,H_2,\cdots,H_{n-1},x_{n-1})\over\partial(x_1,x_2,\cdots,x_n)}
$$
is true. Similarly we can derive
\begin{eqnarray*}
{\partial(H_1,H_2,\cdots,H_{n-1},x_{n})\over\partial(x_1,x_2,\cdots,x_n)}
&=&
\left|\begin{array}{cc}\displaystyle{{\partial H_{n-1}\over\partial x_{n-1}}}&\displaystyle{{\partial H_{n-1}\over\partial x_{n}}}\\
\displaystyle{{\partial x_{n}\over\partial x_{n-1}}}&1\\\end{array}\right|\\
&=&
\det J-{\partial H_{n-1}\over\partial x_{n}}{\partial x_{n}\over\partial x_{n-1}}\\
&=&
\det J.
\end{eqnarray*}
The last equality follows to the fact $\partial x_{n}/\partial x_{n-1}=0$. 

These results, together with the fact $dx_j/dx_n=0,\ j=1,2,\cdots, n-2$, can be summaried to the formula
\begin{equation}
{dx_j\over dx_n}={1\over \det J}{\partial(H_1,H_2,\cdots, H_{n-1},x_j)\over \partial(x_1,x_2,\cdots,x_n)},\qquad j=1,2,\cdots,n.
\end{equation}
It is now straightforward to show (\ref{Nambu-Hamilton eq})
\begin{eqnarray*}
{dX_j\over dx_n}&=&\sum_{k=1}^n{\partial X_j\over\partial x_k}{dx_k\over dx_n}={1\over \det J}\sum_{k=1}^n{\partial X_j\over\partial x_k}{\partial(H_1,H_2,\cdots, H_{n-1},x_k)\over \partial(x_1,x_2,\cdots,x_n)}\\
&=&
\sum_{k=1}^n{\partial X_j\over\partial x_k}{\partial(H_1,H_2,\cdots, H_{n-1},x_k)\over \partial(X_1,X_2,\cdots,X_n)}\\
&=&{\partial(H_1,H_2,\cdots, H_{n-1},X_j)\over \partial(X_1,X_2,\cdots,X_n)},\qquad j=1,2,\cdots,n.\\
\end{eqnarray*}
To derive the last line we used the identity (\ref{identity}).

\section{The map-flow correspondence}

Having established our proposition we would like to clarify the meaning of our results. In summary the proposition provides a view to see an arbitrary differentiable map as a dynamical flow. When $(x_1,x_2,\cdots,x_{n-2})$ are given as functions of $(X_1,X_2,\cdots,X_n)$ and the Jacobian of the map is known, we can reproduce whole of the map $(x_1,x_2,\cdots,x_{n})\rightarrow (X_1,X_2,\cdots,X_n)$ by solving the Nambu-Hamilton equations under proper initial conditions. Therefore the Nambu-Hamiltonian flow is equivalent with the map in the sense that they are different view of one relation. We will present various examples of such maps in the next section.

After $m$ steps of the map $f$ we obtain the map
\begin{equation}
\mbox{\boldmath$x$}=(x_1,x_2,\cdots,x_n)\ \rightarrow\ \mbox{\boldmath$X$}=(f^{(m)}_1(\mbox{\boldmath$x$}),f^{(m)}_2(\mbox{\boldmath$x$}),\cdots, f^{(m)}_n(\mbox{\boldmath$x$})).
\label{m maps}
\end{equation}
We can apply all of the arguments of the previous section to this map. For instance the Hamiltonians $H_1, H_2,\cdots, H_{n-2}$ are given by (\ref{H's}), {\it i.e.}, $H_j=x_j=f_j^{(-m)}(\mbox{\boldmath$X$}),\ j=1,2,\cdots, n-2$. Only the difference from the single map is that the expressions of \mbox{\boldmath$x$} become complicated in the case of $m$ steps when they are written explicitly in terms of \mbox{\boldmath$X$}. If $J^{(k)}$ is the Jacobi matrix from the variable $\mbox{\boldmath$x$}^{(k)}=f^{(k)}(\mbox{\boldmath$x$})$ to $\mbox{\boldmath$x$}^{(k+1)}=f(\mbox{\boldmath$x$}^{(k)})$, the Jacobi matrix $J$ of the map (\ref{m maps}) is a product of $J^{(1)}, J^{(2)},\cdots, J^{(m)}$. Hence we have
$$
\det J=\det J^{(1)}\det J^{(2)}\cdots \det J^{(m)}.
$$
From this information the Hamiltonian $H_{n-1}(\mbox{\boldmath$X$})$ can be also calculated according to (\ref{H's}). This observation enables us to see the map after $m$ steps also as a dynamical flow governed by the Nambu-Hamilton equations. Namely it is a flow of $\mbox{\boldmath$X$}$ such that $n-1$ Hamiltonians $H_1(\mbox{\boldmath$X$}),H_2(\mbox{\boldmath$X$}),\cdots$ $,H_{n-1}(\mbox{\boldmath$X$})$ remain unchanged. It will become quite simple if the Jacobian is a constant.

In the study of dynamical systems in one dimension it is well known that the nature of an orbit of a map is determined by local properties of the map function near the fixed and/or periodic points of the map. There have been, however, known a little about the case of higher dimensional maps\cite{Devaney}. The main difficulty of the analysis of higher dimensional maps lies in its complexity. There is no systematic method, for example, to study local properties of the map when initial values are varied. This must be contrasted with the case of one dimensional map. Various notions well studied in the one dimensional case, such as chaos attractors, Julia sets etc., are not clear in the higher dimensional cases.

We would like to emphasize here that the Nambu-Hamilton equations can be considered as equations which determine local properties of the map when one of the initial values is varied. Let us rewrite the equation (\ref{Nambu-Hamilton eq}) as
\begin{equation}
{dX_j\over dx_n}=(-1)^{n-j}{\partial(H_1,H_2,\cdots, H_{n-1})\over\partial(X_1,X_2,\cdots,\hat X_j,\cdots,X_n)},
\label{NH eq}
\end{equation}
where $\hat X_j$ in this expression means that $X_j$ is missing from $(X_1,X_2,\cdots,X_n)$. This formula shows that the variation of $X_j$ is determined by the Jacobian of other variables. The set of equations (\ref{NH eq}) determine a one parameter orbit in the $n$ dimensional space of $(X_1,X_2,\cdots,X_n)$ such that $H_1(\mbox{\boldmath$X$}),H_2(\mbox{\boldmath$X$}),\cdots,H_{n-1}(\mbox{\boldmath$X$})$ are fixed. When we study behaviour of the map along this orbit, other variables $x_j\ne x_n$ are irrelevant. Moreover the set of equations (\ref{NH eq}) are sufficient to reconstruct the map if the initial conditions are chosen properly. Therefore the Nambu-Hamilton equations should provide a useful tool to analyse the map systematically.

\section{Examples}

We will show, in this section, some examples of dynamical systems which exhibit the map-flow correspondence.

\subsection{Hermite polynomials}

Let us start from two dimensional maps. We consider first the recurrence formula for Hermite polynomials ${\cal H}_k(x)$
\begin{equation}
{\cal H}_{k+1}-x{\cal H}_{k}+k{\cal H}_{k-1}=0,\quad k=1,2,3,\cdots.
\label{Hermite recurrence formula}
\end{equation}
If we define new variables 
\begin{equation}
x^{(k)}=x,\qquad y^{(k)}={\cal H}_k/{\cal H}_{k-1}
\label{y=H/H}
\end{equation}
we obtain a map
\begin{equation}
\left(\begin{array}{c} x^{(k+1)} \cr y^{(k+1)} \cr \end{array}\right)=
\left(\begin{array}{c}x^{(k)}\cr x^{(k)}-\displaystyle{{k\over y^{(k)}}} \cr\end{array}\right),\qquad k=1,2,\cdots .
\label{Hermite map}
\end{equation}
The Jacobi matrix of this map is
$$
J^{(k)}=\left(\begin{array}{cc}1&0\cr 1&k/(y^{(k)})^2\cr\end{array}\right).
$$
After $m$ steps of the map, $(x,y):=(x^{(1)},y^{(1)})\rightarrow (X,Y):=(x^{(m)},y^{(m)})$, the Jacobian is given by
\begin{equation}
\det J:={(m-1)!\over \left(y^{(m-1)}y^{(m-2)}\cdots y^{(1)}\right)^2},
\label{det J^m}
\end{equation}
and the Hamiltonian, according to (\ref{Hamiltonian}), turns out to be
\begin{eqnarray}
m=2:\quad &&H(X,Y)=X(X-Y)^2\nonumber\\
m=3:\quad &&H(X,Y)={(2-X(X-Y))^2\over Y-X},\qquad etc.
\label{H(XY) of Hermite}
\end{eqnarray}
These data are sufficient to derive the Hamilton equations of motion (\ref{Hamilton's equations}) associated to the map (\ref{Hermite map}). In the case of $m=2$, for instance, they are 
\begin{eqnarray}
{dX\over dy}=2X(X-Y),\qquad {dY\over dy}=(Y-X)^2+2X(X-Y).
\label{Hamilton eq of Hermite m=2}
\end{eqnarray}
The equations (\ref{Hamilton eq of Hermite m=2}) can be solved easily and we find
$$
X=x,\qquad Y=x-{1\over y},
$$
when the initial values are chosen to satisfy 
$$
x=cy^2.
$$
Since the parameter $c$ is arbitrary the last condition does not restrict the initial values $(x,y)$. Thus we reproduced successfully the map from which we started. 

In the case of $m=3$ the equations of motion are
\begin{eqnarray}
{dX\over dy}&=&-X^2+{4\over (X-Y)^2},\nonumber\\
{dY\over dy}&=&4-X^2+{4\over(X-Y)^2}+2X(Y-X),
\label{Hamilton eq of Hermite}
\end{eqnarray}
which are solved by the map
$$
X=x,\qquad Y=x-{2y\over xy-1}
$$
only if the initial values are constrained by 
$$
x={c\over y^2}+{1\over y},\qquad c:{\rm const.}
$$

We could derive the Hamilton equations in which $x$ plays the role of the time variable instead of $y$. In this case, we first notice the relation
$$
{\partial Y\over\partial x}=\det J
$$
which will be verified by comparing the expressions in both sides. The Hamiltonian is thus given by
$$
H(X,Y)=\int^y\det J\ dy=Y+h(x).
$$
Here we introduced an arbitrary function $h(x)$ of $x$ , which does not affect the Hamilton equations derived from the map. (We have not used this possibility in (\ref{H(XY) of Hermite}) because it was irrelevant for the purpose of reproducing the map.) Using the nature of the map $X=x$, the Hamilton equations
$$
{dX\over dx}={\partial H\over\partial Y},\qquad {dY\over dx}=-{\partial H\over\partial X}
$$
associated to this system are 
\begin{equation}
{dX\over dx}=1,\qquad {dY\over dx}=-{dh(X)\over dX}.
\label{Hamilton eq for Hermite pol}
\end{equation}

Solving these equations is rather trivial in this case and we obtain
\begin{equation}
X=x\qquad Y(x)=-h(x).
\label{Y=-h(x)}
\end{equation}
The map-flow correspondence is achieved if $h(x)+x$ is given by the following continued fraction
$$
h(x)+x={m-1\over y^{(m-1)}(x)}={m-1\over x} { \atop -}{m-2\over x}{ \atop -}\cdots{ \atop -} {2\over x}{ \atop -} {1\over x}.
$$

We note that the map (\ref{Y=-h(x)}) is nothing but the recurrence formula (\ref{Hermite recurrence formula}) of the Hermite polynomials. On the other hand, if we substitute (\ref{y=H/H}) into the Hamilton equation (\ref{Hamilton eq for Hermite pol}) and denote ${\cal H}'_{m}=d{\cal H}_{m}/dx$, we obtain a differential-difference equation
\begin{eqnarray*}
{\cal H}_{m}{\cal H}'_{m+1}-{\cal H}_{m+1}{\cal H}'_{m}-{\cal H}^2_{m}+m{\cal H}_{m}{\cal H}'_{m-1}-m{\cal H}_{m-1}{\cal H}'_{m}=0,
\end{eqnarray*}
which can be also written as 
\begin{eqnarray*}
&&{\cal H}_{m}\left({\cal H}'_{m+1}-(m+1){\cal H}_{m}\right)
-{\cal H}_{m+1}\left({\cal H}'_{m}-m{\cal H}_{m-1}\right)\\
&&+m\left({\cal H}'_{m-1}-x{\cal H}_{m-1}+{\cal H}_{m}\right){\cal H}_{m}
-m\left({\cal H}'_{m}-x{\cal H}_{m}+{\cal H}_{m+1}\right){\cal H}_{m-1}=0
\end{eqnarray*}
This is solved if 
$$
{\cal H}_{m+1}=x{\cal H}_{m}-{\cal H}'_{m},\qquad {\cal H}_{m-1}={1\over m}{\cal H}'_{m}
$$
or equivalently the Hermite differential equation
\begin{equation}
{d^2{\cal H}_{m}\over dx^2}-x{d{\cal H}_{m}\over dx}+m{\cal H}_{m}=0
\label{Hermite equation}
\end{equation}
holds.

\subsection{H\'enon map}

Next we study the H\'enon map\cite{Henon} as another example of two dimensional map, but not integrable\cite{FM}. This is a canonical form of the two dimensional polynomial diffeomorphism. For arbitrary constants $b,\ c$ it is given by
\begin{equation}
\left(\begin{array}{c}X\cr Y\cr\end{array}\right)
=
\left(\begin{array}{c}y\cr y^2-bx+c\cr\end{array}\right),\qquad
\left(\begin{array}{c}x\cr y\cr\end{array}\right)
=
\left(\begin{array}{c}(X^2-Y+c)/b\cr X \cr\end{array}\right),
\label{Henon}
\end{equation}
\[
J=\left(\begin{array}{cc}0&1\cr -b&2y \cr\end{array}\right),\qquad
J^{-1}=\left(\begin{array}{cc}2X/b&-1/b\cr 1&0 \cr\end{array}\right).
\label{J,J^-1}
\]
Since $\det J=b$, we have simply $H=bx+ const$, which can be also written as
\begin{equation}
H(X,Y)=X^2-Y+c.
\label{Henon H(X,Y)}
\end{equation}
If we repeat the map we find different forms of the Hamiltonian
\begin{eqnarray*}
m=3:\quad &&H(X,Y)=(X^2-Y+c)^2/b-Xb+cb\\
m=4:\quad &&H(X,Y)=((X^2-Y+c)^2/b^2-X+c)^2-(X^2-Y+c)b+cb^2,\qquad etc.
\end{eqnarray*}

Now we like to know, when the Hamiltonian is given by (\ref{Henon H(X,Y)}) and the Jacobian satisfies $\det J=b$, if the map is recovered. The Hamilton equations (\ref{Hamilton's equations})
$$
{dX\over dy}=1,\qquad {dY\over dy}=2X
$$
are solved by
$$
X(y)=y+\alpha,\quad Y(y)=y^2 +2\alpha y+\beta.
$$
The constants $\alpha,\ \beta$ can depend on $x$. Imposing the condition
$$
\det J=\det\left(\begin{array}{cc}\displaystyle{{d\alpha\over dx}}&1 \cr 2y\displaystyle{{d\alpha\over dx}}+\displaystyle{{d\beta\over dx}}&2(y+\alpha) \cr\end{array}\right)={d\over dx}(\alpha^2-\beta)=b
$$
we obtain
$$
\alpha^2-\beta=bx-c,\qquad c \ :{\rm constant}.
$$
The H\'enon map (\ref{Henon}) corresponds to the case of $\alpha=0$ and $\beta=-bx+c$. When $b=1$, the map is an area preserving map.

In contrast to the previous example of Hermite polynomials the H\'enon map is not integrable, hence there is no analytic expression for the Hamiltonian. This fact obstructs us from deriving the Hamilton equations in a systematic way.


\subsection{3-point KdV map}

The purpose of this subsection is to illustrate the Nambu-Hamiltonian flow associated with the higher dimensional maps. To simplify the problem, however, we consider only three dimensional cases. In particular we study two examples, {\it i.e.}, the discrete KdV equation\cite{Hirota} and the $q$-difference Painlev\'e IV map. 

The KdV map
\begin{equation}
\frac{1}{x_{j}}-\frac{1}{X_{j}}=X_{j+1}-x_{j-1},\qquad j\in \mbox{\boldmath{Z}}.
\end{equation}
is known completely integrable. The general solutions are given in terms of the $\tau$ function. It is a generalization of the Korteweg-de Vries equation to discrete time and discrete space $j$. To be specific we restrict our argument to three point KdV map $(x_{1}, x_{2}, x_{3})=
(x,y,z)$ with the periodic boundary conditions $x_{j}=x_{j+3}$. Then the equations become
\begin{eqnarray}
\frac{1}{x}-\frac{1}{X}&=&Y-z,\nonumber\\
\frac{1}{y}-\frac{1}{Y}&=&Z-x,\label{Discrete 3-point KdV}\\
\frac{1}{z}-\frac{1}{Z}&=&X-y.\nonumber
\end{eqnarray}
The integrability of this map is obvious since there are two constants of the map, which we denote by $u$ and $v$ :
\[
u=\frac{1}{xyz}+xyz,\qquad v=\frac{1}{x}+\frac{1}{y}+\frac{1}{z}+x+y+z.
\]

Solving (\ref{Discrete 3-point KdV}) for time evolution, we get two types of solutions\cite{NSSY}. The map which we discuss in this paper is the one given by
\begin{equation}
\left(\begin{array}{c}X\cr Y\cr Z\cr\end{array}\right)
=
\left(\begin{array}{c}x\displaystyle{{1+xy+xy^2z\over 1+zx+x^2yz}}\cr
\quad\cr
y\displaystyle{{1+yz+xyz^2\over 1+xy+xy^2z}}\cr
\quad\cr
z\displaystyle{{1+zx+x^2yz\over 1+yz+xyz^2}}\cr\end{array}\right),
\label{KdV map (x,y,z) to (X,Y,Z)}
\end{equation}
or conversely
\begin{equation}
\left(\begin{array}{c}x\cr y\cr z\cr\end{array}\right)
=
\left(\begin{array}{c}X\displaystyle{{1+ZX+XYZ^2\over 1+XY+X^2YZ}}\cr
\quad\cr
Y\displaystyle{{1+XY+X^2YZ\over 1+YZ+XY^2Z}}\cr
\quad\cr
Z\displaystyle{{1+YZ+XY^2Z\over 1+ZX+XYZ^2}}\cr\end{array}\right).
\label{KdV map (X,Y,Z) to (x,y,z)}
\end{equation}

\subsubsection{Three dimensional view}

We immediately find the Jacobian $\det J$ of the three dimensional map $(x,y,z)\rightarrow (X,Y,Z)$ being 1, hence the volume is preserved under the map. According to the prescription of the previous section the Nambu Hamiltonians are
$$
H_1=x+c_1,\quad H_2=y+c_2,\qquad c_1,\ c_2 \ :\ {\rm constants}.
$$
The Nambu equations are given explicitly by
\begin{eqnarray*}
{dX\over dz}&=&{\partial(H_1,H_2)\over\partial(Y,Z)}=-{X^2(1-Y^2+2YZ+Y^2Z^2)\over(1+YZ+XY^2Z)^2}\\
{dY\over dz}&=&{\partial(H_1,H_2)\over\partial(Z,X)}={Y^2(1+2ZX+2XYZ^2+X^2Y^2Z^2)\over(1+YZ+XY^2Z)^2}\\
{dZ\over dz}&=&{\partial(H_1,H_2)\over\partial(X,Y)}={1+2ZX+2XYZ^2+X^2Y^2Z^2 \over (1+YZ+XY^2Z)^2}\\
\end{eqnarray*}
If we take into account the fact that $x,y$ are invariants of the Nambu flow, we see that the equations are solved by $(X,Y,Z)$ of (\ref{KdV map (x,y,z) to (X,Y,Z)}).

\subsubsection{Two dimensional view}

We now like to see this dynamical system from another point of view. First we note that the map ($\ref{KdV map (x,y,z) to (X,Y,Z)}$) and ($\ref{KdV map (X,Y,Z) to (x,y,z)}$) have two independent constants of map which we denote by $r$ and $s$ :
\[
r=xyz,\qquad s=(1+xy)(1+yz)(1+zx).
\]
The previous constants $u$ and $v$ can be fixed if $r$ and $s$ are fixed. Using these constants we can reduce the number of independent variables. For instance, by using $r$ we can eliminate $z$ from the map and write (\ref{KdV map (x,y,z) to (X,Y,Z)}) as
\begin{eqnarray}
\left(\begin{array}{c}X\cr Y\cr\end{array}\right)
&=&
\left(\begin{array}{c}\displaystyle{{xy(1+ry+xy) \over r + y + rxy}}\cr
\quad\cr
\displaystyle{{r^2+ry+xy \over x(1+ry+xy)}}\cr\end{array}\right),
\label{KdV raising map}
\\
&&\qquad \nonumber\\
&&\qquad \nonumber\\
\left(\begin{array}{c}x\cr y\cr\end{array}\right)
&=&
\left(\begin{array}{c}\displaystyle{{r^2+rX+XY \over Y(1+rX+XY)}}\cr
\quad\cr
\displaystyle{{XY(1+rX+XY) \over r+X+rXY}}\cr\end{array}\right).
\label{KdV lowering map}
\end{eqnarray}
The result of the proposition 1 is applicable to this problem.

The Jacobian of the transformation turns out to be
\begin{equation}
\det J={r^2+ry+xy\over x(r+y+rxy)},
\label{J_jj+1}
\end{equation}
from which we can calculate the Hamiltonian
\begin{eqnarray}
H&=&r\ln x+\left({1\over r}-r\right)\ln(r+y+rxy)
\nonumber\\
&=&r\ln {r+X+rXY\over Y(1+rX+XY)^2}+
{1\over r}\ln{(r^2+rX+XY)(1+rX+XY)\over r+X+rXY}
\label{H}
\end{eqnarray}

Contrary to the three dimensional view the variable $y$ appears in the expression of the Hamiltonian explicitly. This $y$ dependence can be eliminated if $x$ also depends on $y$. That is, 
$$
{dH\over  dy}={\partial H\over \partial y}+{\partial H\over \partial x}{dx\over dy}=0
$$
if 
\begin{equation}
{dx\over dy}={x(r^2-1)(1+rx)\over r(r^2+ry+xy)}
\label{dx/dy}
\end{equation}
is satisfied.
From (\ref{KdV raising map}) and (\ref{dx/dy}) we obtain
\begin{eqnarray}
{dX\over dy}&=&{\partial X\over \partial y}+{\partial X\over \partial x}{dx\over dy}\nonumber\\
&=&{(r^3x^2y^2+4r^2y^2x-2xy^2+2r^3xy+2r^3y^2-ry^2-y+2r^4y+r^2y+r^3)x\over (r+y+rxy)(r^2+ry+xy)r}
\label{dX/dy}\\
{dY\over dy}&=&{\partial Y\over \partial y}+{\partial Y\over \partial x}{dx\over dy}\nonumber\\
&=&{(1-r^2)(x^2y+2rxy+2r^2x+r^2y+r^3+r)(r+y+rxy)\over x(1+ry+xy)^2(r^2+ry+xy)r}.
\label{dY/dy}
\end{eqnarray}

On the other hand we can calculate $- \partial H/\partial Y$ and $\partial H/\partial X$ from (\ref{H}) and see that they coincide with (\ref{dX/dy}) and (\ref{dY/dy}), respectively. Thus we derived the Hamilton equations of motion (\ref{Hamilton's equations}). In this flow the Hamiltonian $H$ of (\ref{H}) and $r=XYZ$ remain invariant.

This example clearly shows us that the three dimensional Nambu equation is a proper way to characterize the 3d KdV flow. The same flow can be formulated in terms of two dimensional Hamilton equations, but we must pay the cost of cumbersome description.

\subsection{$q$-difference Painlev\'e IV map}

Another example of three dimensional map, which we are going to present here, is the $q$-difference Painlev\'e IV map. The continuous version of this map is known as the Painlev\'e IV equation (P$_{IV}$). It is completely integrable although there is no conserved quantity. A $q$-difference version of P$_{IV}$ has been introduced recently and discussed from the view point of its symmetry\cite{KNY}. This symmetry generates B\"acklund transformations and enables one to obtain some of special solutions. Nevertheless the integrability of the map itself is not yet obvious.

The map we consider is given by
\begin{equation}
\left(\begin{array}{c}X\cr Y\cr Z\cr\end{array}\right)
=
\left(\begin{array}{c}aby\displaystyle{{1+cz+cazx\over 1+ax+abxy}}\cr
\quad\cr
bcz\displaystyle{{1+ax+abxy\over 1+by+bcyz}}\cr
\quad\cr
cax\displaystyle{{1+by+bcyz\over 1+cz+cazx}}\cr\end{array}\right),\quad
\left(\begin{array}{c}x\cr y\cr z\cr\end{array}\right)
=
\left(\begin{array}{c}\displaystyle{{Z\over ca}}\displaystyle{{1+Y/b+YX/ba\over 1+X/a+XZ/ac}}\cr
\quad\cr
\displaystyle{{X\over ab}}\displaystyle{{1+Z/c+ZY/cb\over 1+Y/b+YX/ba}}\cr
\quad\cr
\displaystyle{{Y\over bc}}\displaystyle{{1+X/a+XZ/ac\over 1+Z/c+ZY/cb}}\cr\end{array}\right).
\label{dPIV map}
\end{equation}
Here $a,b,c$ are some constants. If we have chosen the parameters such that $a=b=c=1$ or $a=b=c=-1$, there exist constants of the map
\[
r=xyz,\qquad s=(1+ax)(1+by)(1+cz)
\]
and hence the map becomes completely integrable. They are nothing but special cases of the 3-point Lotka-Volterra map discussed in \cite{NSSY}. Otherwise the map is called $q$-difference Painlev\'e IV \cite{KNY}.

This map is three dimensional. The Jacobian of the map is $q^2=(abc)^2$ in this case. The Nambu Hamiltonians are now given by
$$
H_1=x+c_1\quad H_2=y+c_2.
$$
The Nambu equations are
\begin{eqnarray*}
{dX\over
dz}&=&{\partial(H_1,H_2)\over\partial(Y,Z)}={q(X/a)(1+X/a)(1+Z/c+YZ/bc)\over
b(1+Y/b+XY/ab)(1+X/a+ZX/ca)},\\
{dY\over
dz}&=&{\partial(H_1,H_2)\over\partial(Z,X)}={q(1+Y/b)(1+Z/c+YZ/bc)\over
a(1+Y/b+XY/ab)(1+X/a+ZX/ca)},\\
{dZ\over dz}&=&{\partial(H_1,H_2)\over\partial(X,Y)}=-{c^2(Z/c)(1+Z/c+YZ/bc) \over (1+X/a+ZX/ca)(1+Y/b+XY/ab)},\\
\end{eqnarray*}
which are solved by $(X,Y,Z)$ of (\ref{dPIV map}).


\section{Concluding remarks}

We have shown a correspondence between a map and a Nambu-Hamiltonian flow. Examples presented in this paper are those of lower dimensional map. There is no difficulty to study higher dimensional maps if we do not hesitate lengthy-formulae. As we discussed in \S3, our formulation can apply to every step of a sequence of the map. We would like to report some results elsewhere which can be obtained from the study in this direction.

It is interesting that, by generalizing the map from two dimensional to higher, we have obtained the Nambu dynamics quite naturally. The Nambu equation, which was proposed 30 years ago\cite{Nambu}, was reformulated in 1993 by Takhtajan\cite{Takhtajan} from geometrical point of view. Quantization of the Nambu system has been attempted within the framework of deformation quantization\cite{DFST}. Recently it has attracted attention\cite{ALMY} under the hope that it might provide a useful tool for the space-time quantization.

The map-flow correspondence, which we discussed in this paper, is common to all differentiable maps which have Jacobian. Interpretation of the correspondence can be different depending on the problem. As we presented in \S 4, the Hermite polynomials can be obtained by solving either the map equations (\ref{Hermite map}) or Hamilton equations (\ref{Hamilton eq of Hermite}). They are equivalent to the recurrence formula (\ref{Hermite recurrence formula}) and differential equations (\ref{Hermite equation}), respectively. In this example a sequence of the map has a meaning of increase of the degree of polynomials. If we are considering a system of quantum oscillators, the map transforms an orbit to another of different quantum number, whereas the Hamilton equations determine orbits. This particular interpretation can be shared by all of orthogonal polynomials, since the Hermite polynomials are not exceptional.

Our consideration of this paper is motivated by the observation found by 
one of the authors (A.S.) during the study of quantized H\'enon map to 
which the exact WKB method is applied\cite{ShudoIkeda}. The exact 
WKB analysis of quantized H\'enon map is just to examine the Stokes 
geometry of quantum propagator which is represented by a multiple integral. 
However, since the higher dimensional exact WKB method is mainly 
developed for differential equations\cite{AKT}, so one has to map the problem to 
that of the differential equation. In order to deal with the Stokes 
geometry of higher order differential equations, it is essential to 
consider the bicharacteristic equation induced by the Borel transform of 
differential equation under consideration\cite{AKT}.  A prototype of  
Hamiltonian formalism, which is indeed generalized in this paper, 
appears as a bicharacteristic equation in which the initial value of 
the original H\'enon map plays a role of the time variable.

Some of integrable systems have been studied within the framework of Nambu mechanics\cite{Guha}. Since we are interested in the correspondence between a discrete map and its associated flow we restricted our study to discrete integrable systems. Besides the KdV and Painlev\'e IV maps, which we presented in this paper, there are known many discrete systems which are completely integrable. We have studied some of them, such as discrete Lotka-Volterra system, discrete time Toda lattice, discrete KP hierarchies, and found that the Jacobian of all of them is one. The map of such a system can be interpreted as a B\"acklund transformation from one equation to another which preserves integrability.

We could discuss a theory of renormalization group. Renormalizing a mass $m^{(1)}$ and a coupling constant $g^{(1)}$ at a renormalization point $\mu^{(1)}$, we obtain a map $(m^{(1)},g^{(1)},\mu^{(1)})\rightarrow (m^{(2)},g^{(2)},\mu^{(2)})$. If the theory is renormalizable this procedure forms a group and the sequence of the map converges. Instead of repeating the map, however, there exists a convenient way to get useful information of the renormalized theory. Namely the scale invariance of the theory enables one to derive a differential equation, called the Gell-Mann Low equation, in which the initial value of $\mu$ becomes an independent variable. Solving this equation we can predict about phase properties of the theory. We would like to discuss on this topics from our point of view elsewhere.

\noindent
{\bf Acknowledgement} The authors would like to thank Dr. N.Saitoh for careful reading of the manuscript and useful comments.

\vfill\eject
\section*{Appendix :
 One dimensional map}

We could consider a simpler case of the map, {\it i.e.}, a sequence of one dimensional differentiable map:
\[
q_k\rightarrow q_{k+1},\qquad k=0,1,2,\cdots, m-1
\]
and its inverse $q_k\rightarrow q_{k-1}$. This, however, does not correspond to any interesting system as far as the map 
is determined by the nearest neighbours.

An interesting map is obtained if a change of $q_k$ is determined by the change of two neighbours. To be specific let us 
consider a map defined by
\begin{equation}
q_k=\alpha(q_{k+1})-\beta(q_{k+2}),\qquad k=0,1,2,\cdots, m-2
\label{q_k map}
\end{equation}
with some differentiable functions $\alpha$ and $\beta$. We assume $\beta$ is invertible, so that the inverse map of 
(\ref{q_k map}) exists:
\begin{equation}
q_k=\beta^{-1}(\alpha(q_{k-1})-q_{k-2}),\qquad k=2,3,\cdots, m.
\end{equation}
Writing them in derivative forms they are
\begin{eqnarray}
\partial_{k}&=&a_{k}\partial_{k-1}-b_{k-1}\partial_{k-2}\label{q_k=q_k-1}\\
\partial_{k}&=&\bar a_{k}\partial_{k+1}-c_{k+1}\partial_{k+2}
\label{q_k=q_k+1}
\end{eqnarray}
where we have put $\partial_k:=\partial/\partial q_k$ and
\[
a_k:=\left.{d\alpha(x)\over dx}\right|_{x=q_{k}},\qquad 
b_{k}:=\left.{d\beta(x)\over dx}\right|_{x=q_{k+1}}
\]
\begin{equation}
\bar a_k:=\left.{d\alpha(x)\over dx}\right|_{x=q_{k}}\cdot\left.{d\beta^{-1}(x)\over dx}\right|_{x=\beta(q_{k+1})},\qquad
c_{k}:=\left.{d\beta^{-1}(x)\over dx}\right|_{x=\beta(q_{k+1})}.
\label{bar a=ac}
\end{equation}
Since $q_{-1}$ and $q_{m+2}$ are not variables, we assume
\[
b_0=c_{m+1}=0.
\]

By iteration of (\ref{q_k=q_k-1}) we can see a responce of $q_l$ under the variation of previous variables, say 
$(q_{k-1}, q_{k-2}$):
\begin{eqnarray}
\partial_l
&=&(a_{l}a_{l-1}-b_{l-1})\partial_{l-2}-a_{l}b_{l-2}\partial_{l-3}\ =\ \cdots\nonumber\\
&=&
A_{k,l}\partial_{k-1}-b_{k-1}A_{k+1,l}\partial_{k-2}
\label{partial_l}
\end{eqnarray}
where
\[
A_{k,l}:=\det\left(\begin{array}{ccccccc}
a_{k}&b_k&0&0&0&\cdots&0\cr
1&a_{k+1}&b_{k+1}&0&0&\cdots&0\cr
0&1&a_{k+2}&b_{k+2}&0&\cdots&0\cr
0&0&\cdots&&&\cdots&0\cr
&&&\cdots&&&\cr
0&0&\cdots&0&1&a_{l-1}&b_{l-1}\cr
0&0&\cdots&0&0&1&a_{l}\cr
\end{array}\right).
\]
Similarly from (\ref{q_k=q_k+1}) we obtain
\begin{eqnarray}
\partial_k&=&
(\bar a_{k}a_{k+1}-c_{k+1})\partial_{k+2}-\bar a_{k}c_{k+2}\partial_{k+3}\ =\ \cdots\nonumber\\
&=&
\bar A_{k,l}\partial_{l+1}-c_{l+1}\bar A_{k,l-1}\partial_{l+2}
\label{partial_k}
\end{eqnarray}
where
\[
\bar A_{k,l}:=\det\left(\begin{array}{ccccccc}
\bar a_{k}&1&0&0&0&\cdots&0\cr
c_{k+1}&\bar a_{k+1}&1&0&0&\cdots&0\cr
0&c_{k+2}&\bar a_{k+2}&1&0&\cdots&0\cr
0&0&\cdots&&&\cdots&0\cr
&&&\cdots&&&\cr
0&0&\cdots&0&c_{l-1}&\bar a_{l-1}&1\cr
0&0&\cdots&0&0&c_{l}&\bar a_{l}\cr
\end{array}\right).
\]
We note that (\ref{bar a=ac}) shows $\bar a_k=a_kc_k={a_k/b_k}$. Hence $A$ and $\bar A$ are related by
\begin{equation}
\bar A_{k,l}=c_{k}\cdots c_{l}A_{k,l}.
\label{bar A=cccA}
\end{equation}

We want to find a responce of $(q_{m-1}, q_m)$ when the initial value $q_1$ is changed. We define $(q, p)$, instead of 
$q_m(q_1)$ and $q_{m-1}(q_1)$, as dynamical variables which satisfy
\begin{equation}
\partial_m q(q_1)=1,\qquad \partial_{m+1} q(q_1)=0
\end{equation}
and
\begin{equation}
c_m\partial_{m+1}p+\partial_{m-1}p=0.
\label{cP+P=0}
\end{equation}
We mention that, by the definition of the map (\ref{q_k map}), $q_{m+1}$ is not a dynamical variable.

We can prove the following:

\noindent
{\bf Proposition}

{\it Let $H$ be a function of $(q,p)$ which is given by
\begin{equation}
H=\int^{q_0}c_1\cdots c_{m-1}\partial_{m-1}pdq_0.
\label{one-dim Hamiltonian}
\end{equation}
Then Hamilton's equations
\begin{equation}
{dq\over dq_1}={\partial H\over\partial p},\qquad 
{dp\over dq_1}=- {\partial H\over\partial q}
\label{Hamilton eq for q}
\end{equation}
hold.} 

\noindent
{\bf Proof}

To prove this we consider two cases
\begin{eqnarray}
\partial_1&=&\bar A_{1,m-1}\partial_{m}-c_{m}\bar A_{1,m-2}\partial_{m+1}
\label{partial_1=partial_m}\\
\partial_1&=&\bar A_{1,m}\partial_{m+1}-c_{m+1}\bar A_{1,m-1}\partial_{m+2}=\bar A_{1,m}\partial_{m+1}
\label{partial_1=partial_m+1}
\end{eqnarray}
of (\ref{partial_l}), and apply to $q$ and $p$, respectively, to obtain
\begin{eqnarray}
{dq\over dq_1}&=&\bar A_{1,m-1}
\label{dX/dq_1}\\
{dp\over dq_1}&=&\bar A_{1,m}\partial_{m+1}p.
\label{dP/dq_1}
\end{eqnarray}
Similarly we apply 
\begin{eqnarray}
\partial_{m}&=&A_{1,m}\partial_0-b_0A_{2,m}\partial_{-1}=A_{1,m}\partial_0
\label{partial_m}\\
\partial_{m-1}&=&A_{1,m-1}\partial_0-b_0A_{2,m-1}\partial_{-1}=A_{1,m-1}\partial_0
\label{partial_m-1}
\end{eqnarray}
, which are obtained from (\ref{partial_k}), to $H$ and obtain
\begin{eqnarray}
\partial_mH&=&A_{1,m}\partial_0H
\label{partial_mH}\\
\partial_{m-1}H&=&A_{1,m-1}\partial_0H.
\label{partial_m-1H}
\end{eqnarray}

Let us derive the first equation of (\ref{Hamilton eq for q}) from  (\ref{dX/dq_1}) and (\ref{partial_m-1H}). In fact the right hand side of (\ref{partial_m-1H}) can be written, using (\ref{one-dim Hamiltonian}), (\ref{bar A=cccA}) and (\ref{dX/dq_1}) as
\begin{eqnarray*}
{\rm rhs\ of}\ (\ref{partial_m-1H})&=&c_1c_2\cdots c_{m-1}A_{1 m-1}\partial_{m-1}p\\
&=&
\bar A_{1 m-1}\partial_{m-1}p\\
&=&
{dq\over dq_1}\partial_{m-1}p,
\end{eqnarray*}
whereas the left hand side is 
\[
{\rm lhs\ of}\ (\ref{partial_m-1H})=\partial_{m-1}p{\partial H\over\partial p},
\]
from which we obtain the first equation of (\ref{Hamilton eq for q}). The second equation of (\ref{Hamilton eq for q}) is derived by comparing (\ref{dP/dq_1}) and (\ref{partial_mH}). Using (\ref{one-dim Hamiltonian}), (\ref{cP+P=0}) and (\ref{bar A=cccA}),
\begin{eqnarray*}
{\rm rhs\ of}\ (\ref{partial_mH})&=&c_1c_2\cdots c_{m-1}A_{1 m}\partial_{m-1}p\\
&=&
- c_1c_2\cdots c_{m}A_{1 m}\partial_{m+1}p\\
&=&
- \bar A_{1 m}\partial_{m+1}p=- {\rm rhs\ of}\ (\ref{dP/dq_1})
\end{eqnarray*}
hence the second equation of (\ref{Hamilton eq for q}) is proved.

\vglue 0.5cm
It is known that a one dimensional map equivalent with (\ref{Henon}) can be also generated by the action
\[
S={1\over 2}\sum_{k=1}^{m}(q_{k}-q_{k-1})^2+\sum_{k=1}^{m-1}\left({1\over 3}q_k^3 +cq_k\right).
\]
The equations of motion are
\begin{equation}
{\partial S\over\partial q_k}=-q_{k+1}-q_{k-1}+2q_k+q_k^2+c=0,\qquad k=1,2,\cdots, m-1.
\label{eq. motion}
\end{equation}
Comparing with (\ref{q_k map}) we find
\begin{equation}
\alpha(q_k)=q_k^2+2q_k+c,\qquad \beta(q_k)=q_k,
\label{alpha(q_k)=}
\end{equation}
hence
\[
a_k=2q_k+2,\qquad b_k=c_k=1.
\]
If we specify a solution of (\ref{cP+P=0}) for $p$, so that the canonical pair is given by
\begin{equation}
(q,p)=(q_m,q_{m-1}-q_{m+1}),
\label{P}
\end{equation}
the Hamiltonian (\ref{one-dim Hamiltonian}) turns out to be
\[
H=q_0.
\]
We recall that $q_{m+1}$ in (\ref{P}) is not a dynamical variable but a parameter, which we denote as $q_{m+1}=a$. The equation (\ref{q_k map}) enables us to express $H$ in terms of $q_m$. For instance
\begin{eqnarray*}
m=2:\quad&& H(q, p)=(p+a+1)^2-q-1,\\
m=3:\quad&& H(q, p)=((p+a+1)^2-q)^2-1,
\end{eqnarray*}
where we have put $c=0$.

The correspondence between the two sequences (\ref{Henon}) and (\ref{alpha(q_k)=}) of the H\'enon map is established if we relate the variables by
\[
(X,Y)=(q+1,p+a+1).
\]
The generating function of this transformation is 
\[
W(X,p)=(p+a+1)(X-1).
\]

Before closing, we like to mention a small change of our choice of canonical variables from those of ref.\cite{ShudoIkeda}. In the latter the canonical pair of variables is chosen as
\[
(q', p')=(-q_m, q_m-q_{m-1})=(-q, q-p-a),
\]
while the Hamiltonian is given by
\[
H'={\partial S\over \partial q_1}.
\]
These two variables are related by the canonical transformation generated by the following function:
\[
W(q,p',q_1)=\left({1\over 2}q-p'-a\right)q - q_0q_1.
\]
\end{document}